\documentclass[12pt]{article}
\usepackage{graphicx}
\usepackage[cp1251]{inputenc}
\usepackage{rotating}
\begin{document}
 \noindent {\footnotesize\it Astronomy Reports, 2019, Vol. 63, No 11, pp. 932--943}
 \newcommand{\dif}{\textrm{d}}

 \noindent
 \begin{tabular}{llllllllllllllllllllllllllllllllllllllllllllll}
 & & & & & & & & & & & & & & & & & & & & & & & & & & & & & & & & & & & & & \\\hline\hline
 \end{tabular}

 \vskip 0.5cm
  \centerline{\bf\large Features of the Residual Velocity Ellipsoid of Hot Subdwarfs}
  \centerline{\bf\large from the Gaia DR2 Catalog}
 \bigskip
 \bigskip
  \centerline
 {
 V.V. Bobylev and A.T. Bajkova
 }
 \bigskip
 \centerline{\small \it
 Central (Pulkovo) Astronomical Observatory, Russian Academy of Sciences,}
 \centerline{\small \it Pulkovskoe shosse 65, St. Petersburg, 196140 Russia}
 \bigskip
 \bigskip
 \bigskip

 {
{\bf Abstract}---The evolution of the parameters of the residual
velocity ellipsoid of hot subdwarfs (HSDs) with their position
relative to the Galactic plane is traced, using the HSDs selected
by Geier et al. from the Gaia DR2 catalog. Parameters of the
Galactic rotation are determined for two $|z|$ zones. These are
used to estimate the gradient of the circular rotation velocity,
$V_0,$ versus $|z|,$ found to be $\Delta V_0/\Delta |z|=-71\pm7$
km s$^{-1}$ kpc$^{-1}$. The size of the residual velocity
ellipsoid is
$(\sigma_1,\sigma_2,\sigma_3)=(36.1,27.6,22.8)\pm(0.4,0.8,0.6)$
km/s for HSDs at $|z|<0.5$ kpc and
$(\sigma_1,\sigma_2,\sigma_3)=(56.9,55.8,39.7)\pm(0.9,1.1,0.8)$
km/s for HSDs at $|z|\geq0.5$ kpc. When forming the HSD residual
velocities, the Galactic rotation was taken into account using
individual approaches for each $z$ zone. Parameters of the
residual velocity ellipsoids for HSDs located in four
plane-parallel layers are also determined. The size of the
ellipsoid increases with $z,$ and the inclination of the first
axis relative to the Galactic plane also increases. This
inclination is close to zero in zones close to the Galactic plane,
$z\sim\pm0.2$ kpc, and rises to $\mp12\pm4^\circ$ for
$z\sim\pm0.9$ kpc.
  }

\medskip DOI: 10.1134/S1063772919110027

 \section{INTRODUCTION}
In the Hertzsprung–Russell (HR) diagram, hot subdwarfs (sdO, sdB,
sdOB) occupy a compact region at the blue end of the horizontal
branch; thus, they are at late stages in the evolution of fairly
massive stars. Hot subdwarfs (HSDs) are characterized by core
helium burning. Such stars were first identified by Humason and
Zwicky [1] in a study of super-high luminosity blue stars located
near the Galactic North pole. Subsequent spectroscopy revealed
hydrogen deficiencies for many HSDs. Measurements of their
temperatures and surface gravities [2] made it possible to
correctly identify the position of such stars in the HR diagram.

The releases of high-accuracy data from the Gaia space experiment
[3, 4] opened possibilities for large-scale statistical analyses
for different Galactic subsystems. The catalog contains
trigonometric parallaxes and proper motions for about 1.3 billion
stars. The mean uncertainties in the parallaxes of bright stars
$(G<15^m)$ are 0.02--0.04 milliarcseconds (mas), while those for
faint stars $(G=20^m)$ can be up to 0.7 mas. Similarly, the
proper-motion uncertainties range from 0.05 mas/year for bright
stars $(G<15^m)$ to 1.2 mas/year for faint stars $(G=20^m)$ [5].

Iorio and Belokurov [6] used $\sim$230 000 RR Lyrae stars from the
Gaia DR2 catalog to refine the shape of the Milky Way's halo.
Rowell and Kilic [7] analyzed $\sim$79 000 white dwarfs from the
Gaia DR2 catalog, obtained a new estimate of the Sun's peculiar
velocity relative to the Local Standard of Rest, and detected
considerable oscillations of the vertex (the direction of the
first axis of the residual velocity ellipsoid in the $xy$ plane).
Vertical oscillations in the Galactic disk were analyzed using
Gaia DR2 main-sequence stars [8, 9]. Using globular clusters with
proper motions calculated from Gaia DR2 catalog data [10, 11], a
new estimate of the Galactic mass was obtained in [12]. A
considerable number of studies have dealt with refining the
kinematic parameters of the thin disk [13--15], spiral structure
[16, 17], and open clusters [18, 19] using data on young stars
from the Gaia DR2 catalog.

The statistical and kinematic characteristics of HSDs are of
considerable interest: compared to other subsystems of the Galaxy,
they are not numerous and their properties remain poorly studied.
An analysis of comparatively small samples demonstrates that some
have properties corresponding to those of the thin disk, while
others display properties of the thick disk and halo [20--22],
resembling the kinematics of white dwarfs [23]. Methods aimed at
searching for such stars and identifying them in large-scale
surveys were developed [24], and;the first large catalogs of HSDs
containing necessary data (proper motions, parallaxes) for tens of
thousands of HSDs have appeared [25].

Bobylev and Bajkova [26] demonstrated that these stars display
different kinematics at different positions on the celestial
sphere. They considered two samples with different Galactic
latitudes: low-latitude and high-latitude, dictated by the need to
find scale heights. The aim of the present paper is to continue
the study [26], with stars divided into zones according to their
$z$ coordinates, determine parameters of the Galactic rotation in
these zones, and perform a detailed study of the parameters of the
velocity ellipsoid of the samples obtained.

 \section{METHODS}
 \subsection{Galactic Rotation Parameters}
We consider here only stars having measured proper motions. Thus,
two velocities are known from observations: $V_l=4.74r\mu_l\cos b$
and $V_b=4.74r\mu_b,$ directed along Galactic longitude $l$ and
latitude $b$ and expressed in km/s. The coefficient 4.74 is the
ratio between the number of kilometers in an astronomical unit and
the number of seconds in a tropical year; $r=1/\pi$ is the star’s
distance, calculated from its parallax, $\pi.$ The components of
the proper motion, $\mu_l\cos b$ and $\mu_b$ are expressed in
mas/year.

We determined the parameters of the Galactic rotation curve using
equations derived from the formulas of Bottlinger, where we
expanded the angular velocity $\Omega$ into a series, keeping
terms to second power in $r/R_0:$
 \begin{equation}
 \begin{array}{lll}
 V_l= U_\odot\sin l-V_\odot\cos l
 +r\Omega_0\cos b\\
 -(R-R_0)(R_0\cos l-r\cos b)\Omega^\prime_0
 -0.5(R-R_0)^2(R_0\cos l-r\cos b)\Omega^{\prime\prime}_0,
 \label{EQ-2}
 \end{array}
 \end{equation}
 \begin{equation}
 \begin{array}{lll}
 V_b=U_\odot\cos l\sin b + V_\odot\sin l \sin b-W_\odot\cos b\\
  +R_0(R-R_0)\sin l\sin b\Omega^\prime_0
 +0.5R_0(R-R_0)^2\sin l\sin b\Omega^{\prime\prime}_0,
 \label{EQ-3}
 \end{array}
 \end{equation}
where $R$ is the distance between the star and the Galactic
rotation axis (the corresponding cylindrical radius vector):
  \begin{equation}
 R^2=r^2\cos^2 b-2R_0 r\cos b\cos l+R^2_0.
 \end{equation}
$(U,V,W)_\odot$ is the sample’s group velocity, and reflects the
peculiar motion of the Sun relative to the Local Standard of Rest
(LSR), $V_\odot$ contains the component $V_{lag}$ reflecting the
lag of the sample behind the LSR due to the asymmetric drift
effect, $\Omega_0$ is the angular velocity of the Galactic
rotation at the solar distance $R_0,$ $\Omega^{\prime}_0$ and
$\Omega^{\prime\prime}_0$ are the corresponding derivatives of the
angular velocity, and the linear rotation velocity is
$V_0=|R_0\Omega_0|$. The signs of the unknowns were chosen so that
rotations from the $x$ axis to the $y$ axis, from $y$ to $z,$ and
from $z$ to $x$ are positive. Thus, the angular velocity of the
Galactic rotation will be negative here, in contrast to many other
studies in which the Galactic rotation was taken to be positive,
for convenience [27--29].

In the set of conditional equations (1)--(2), we determined the
six unknowns $U_\odot,$ $V_\odot,$ $W_\odot,$ $\Omega_0,$
$\Omega^\prime_0$ and $\Omega^{\prime\prime}_0.$ from a
least-squares solution to these equations. We applied weights
$w_l=S_0/\sqrt {S_0^2+\sigma^2_{V_l}}$ and $w_b=S_0/\sqrt
{S_0^2+\sigma^2_{V_b}},$ where $S_0$ is the ``cosmic'' dispersion
(which we found beforehand for each sample; it is close to the
unit weight uncertainty $\sigma_0$ obtained in the course of a
preliminary solution of the equations); $\sigma_{V_l}$ and
$\sigma_{V_b}$ are the dispersions of the uncertainties of the
corresponding observed velocities. $S_0$ is comparable to the rms
residual $\sigma_0$ (the unit-weight uncertainty) calculated when
solving conditional equations of the form (1)--(2). We searched
for the solution through several iterations using the $3\sigma$
criterion in order to eliminate stars with large residuals.

We should note here several studies aimed at determining the mean
distance between the Sun and the Galactic center using individual
estimates of this parameter performed over the last decade with
independent methods. Examples include $R_0=8.0\pm0.2$ kpc [30],
$R_0=8.4\pm0.4$ kpc [31], and $R_0=8.0\pm0.15$ kpc [32]. We
adopted $R_0=8.0\pm0.15$ kpc in our study.

 \subsection{The Residual Velocity Ellipsoid}
We estimated the dispersion of the stellar residual velocities
using the following well-known method [33]. Let $U,V,W$ be the
velocities along the coordinate axes $x, y,$ and $z.$ Let us
consider the six second-order moments $a, b, c, f, e, d:$
\begin{equation}
 \begin{array}{lll}
 a=\langle U^2\rangle-\langle U^2_\odot\rangle,\\
 b=\langle V^2\rangle-\langle V^2_\odot\rangle,\\
 c=\langle W^2\rangle-\langle W^2_\odot\rangle,\\
 f=\langle VW\rangle-\langle V_\odot W_\odot\rangle,\\
 e=\langle WU\rangle-\langle W_\odot U_\odot\rangle,\\
 d=\langle UV\rangle-\langle U_\odot V_\odot\rangle,
 \label{moments}
 \end{array}
 \end{equation}
which are the coefficients of the equation for the surface
 \begin{equation}
 ax^2+by^2+cz^2+2fyz+2ezx+2dxy=1,
 \end{equation}
and also the components of the symmetric tensor of moments of the
residual velocities:
 \begin{equation}
 \left(\matrix {
  a& d & e\cr
  d& b & f\cr
  e& f & c\cr }\right).
 \label{ff-5}
 \end{equation}
To determine the values in this tensor in the absence of
radial-velocity data, the following three equations are used:
\begin{equation}
 \begin{array}{lll}
 V^2_l= a\sin^2 l+b\cos^2 l\sin^2 l-2d\sin l\cos l,
 \label{EQsigm-2}
 \end{array}
 \end{equation}
\begin{equation}
 \begin{array}{lll}
 V^2_b= a\sin^2 b\cos^2 l+b\sin^2 b\sin^2 l+c\cos^2 b\\
 -2f\cos b\sin b\sin l-2e\cos b\sin b\cos l+2d\sin l\cos l\sin^2 b,
 \label{EQsigm-3}
 \end{array}
 \end{equation}
\begin{equation}
 \begin{array}{lll}
 V_lV_b= a\sin l\cos l\sin b+b\sin l\cos l\sin b\\
 +f\cos l\cos b-e\sin l\cos b+d(\sin^2 l\sin b-\cos^2\sin b),
 \label{EQsigm-4}
 \end{array}
 \end{equation}
for which least-squares solutions were obtained for the six
unknowns $a,b,c,f,e,d$. We then found the eigenvalues of the
tensor (6): $\lambda_{1,2,3}$ by solving the secular equation:
 \begin{equation}
 \left|\matrix
 {
a-\lambda&          d&        e\cr
       d & b-\lambda &        f\cr
       e &          f&c-\lambda\cr
 }
 \right|=0.
 \label{ff-7}
 \end{equation}
This equation’s eigenvalues are inverse squares of the semi-axes
of the ellipsoid of the velocity moments; at the same time, they
are the squares of the semi-axes of the residual velocity
ellipsoid:
 \begin{equation}
 \begin{array}{lll}
 \lambda_1=\sigma^2_1, \lambda_2=\sigma^2_2, \lambda_3=\sigma^2_3,\qquad
 \lambda_1>\lambda_2>\lambda_3.
 \end{array}
 \end{equation}
We found the directions of the main axes of the tensor (10),
$L_{1,2,3}$ and $B_{1,2,3},$ from the relations:
 \begin{equation}
 \tan L_{1,2,3}={{ef-(c-\lambda)d}\over {(b-\lambda)(c-\lambda)-f^2}},
 \label{ff-41}
 \end{equation}
 \begin{equation}
 \tan B_{1,2,3}={{(b-\lambda)e-df}\over{f^2-(b-\lambda)(c-\lambda)}}\cos L_{1,2,3}.
 \label{ff-42}
 \end{equation}
We estimated the uncertainties in $L_{1,2,3}$ and $B_{1,2,3}$
using the scheme:
 \begin{equation}
 \renewcommand{\arraystretch}{2.2}
  \begin{array}{lll}
  \displaystyle
 \varepsilon (L_2)= \varepsilon (L_3)= {{\varepsilon (\overline {UV})}\over{a-b}},\\
  \displaystyle
 \varepsilon (B_2)= \varepsilon (\varphi)={{\varepsilon (\overline {UW})}\over{a-c}},\\
  \displaystyle
 \varepsilon (B_3)= \varepsilon (\psi)= {{\varepsilon (\overline {VW})}\over{b-c}},\\
  \displaystyle
 \varepsilon^2 (L_1)={\varphi^2 \varepsilon^2 (\psi)+\psi^2 \varepsilon^2 (\varphi)\over{(\varphi^2+\psi^2)^2}},\\
  \displaystyle
 \varepsilon^2 (B_1)= {\sin^2 L_1 \varepsilon^2 (\psi)+\cos^2 L_1 \varepsilon^2 (L_1)\over{(\sin^2 L_1+\psi^2)^2}},
 \label{ff-65}
  \end{array}
 \end{equation}
where
 $$
 \varphi=\cot B_1 \cos L_1, \quad \psi=\cot B_1 \sin L_1.
 $$
Three parameters should be computed beforehand: $\overline
{U^2V^2}$, $\overline {U^2W^2}$ and $\overline {V^2W^2},$; then,
 \begin{equation}
 \renewcommand{\arraystretch}{1.6}
  \begin{array}{lll}
  \displaystyle
 \varepsilon^2 (\overline {UV})= (\overline{U^2V^2}-d^2)/n, \\
  \displaystyle
 \varepsilon^2 (\overline {UW})= (\overline {U^2W^2}-e^2)/n, \\
  \displaystyle
 \varepsilon^2 (\overline {VW})= (\overline {V^2W^2}-f^2)/n,
 \label{ff-73}
  \end{array}
 \end{equation}
where $n$ is the number of stars. This method is interesting
because it enables estimation of the uncertainties independently
for each axis. An exception is the directions $L_2$ and $L_3,$
whose uncertainties are the same because they are calculated using
the same formula.

\begin{figure}[t]
{\begin{center}
   \includegraphics[width=0.9\textwidth]{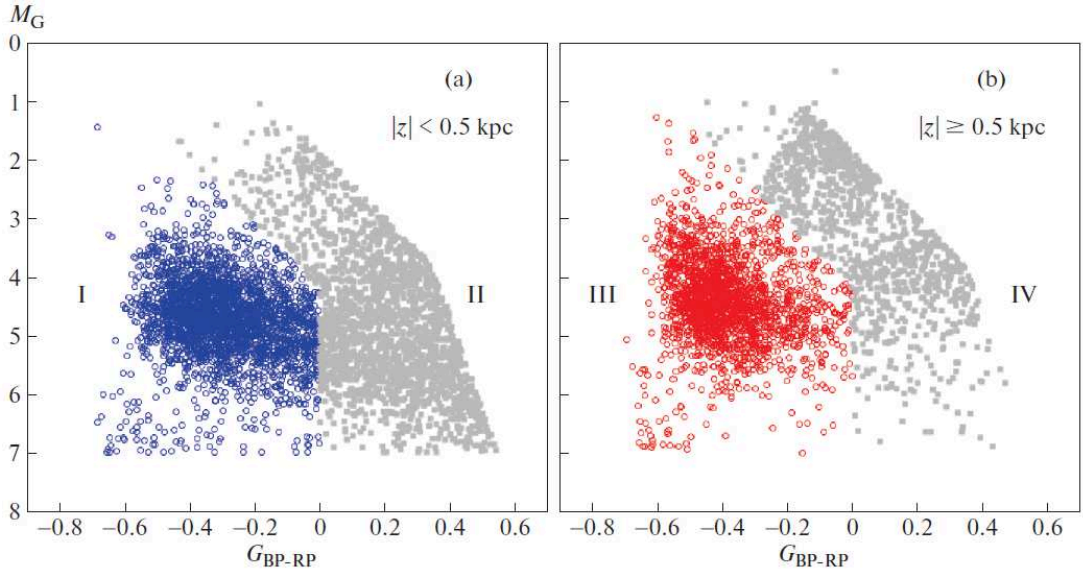}
 \caption{HR diagram for candidate HSDs with relative parallax uncertainties
belower 15\%. See the text for details.
  } \label{f-GR}\end{center}}\end{figure}

 \section*{DATA}
We used the catalog by Geier et al. [25] for our study. This
catalog contains 39 800 candidate HSDs selected from the Gaia DR2
catalog [4, 5], along with satellite and ground-based multiband
photometric and spectroscopic sky surveys, including GALEX, APASS,
SDSS, VISTA, 2MASS, UKIDSS,WISE, etc.

It is supposed that most candidates are HSDs of spectral types O
and B, late-type B stars on the blue horizontal branch, hot stars
on the asymptotic giant branch (post-AGB stars), and central stars
of planetary nebulae. The authors of the catalog believe that the
contamination with cooler stars is about 10\%. The selected HSDs
are distributed over almost the entire celestial sphere. Geier et
al. [25] estimate that, with the exception of the narrow band near
the Galactic plane and the zones containing the Magellanic Clouds,
the catalog is complete out to a distance of 1.5 kpc. The catalog
provides the parallax $\pi$ and two components of the proper
motion, $\mu_\alpha\cos \delta$ and $\mu_\delta,$ for each star.
Radial velocities are not available. Extensive photometric
information is also provided.

In our study, a zero-point correction $\Delta\pi$ was added to the
Gaia DR2 parallaxes, so that the new parallax becomes $\pi+0.050$
mas. The need to take into account this systematic correction
$\Delta\pi=-0.029$ mas (where the minus sign means that the
correction should be added to the Gaia DR2 stellar parallaxes in
order to reduce them to reference values) was first noted by
Lindegren et al. [5] and confirmed by Arenou et al. [34]. Slightly
later, it was demonstrated in [35--39] based on an analysis of
various high-accuracy sources of distance scales that the
correction was about $\Delta\pi=-0.050$ mas, with an uncertainty
of several units in the last digit.

Note that the mean shift of the parallax zero point specifically
for the HSDs is a matter of discussion. Some data indicate that
the correction may be less significant, from about $-0.020$ to
about $-0.030$ mas. This may follow, for example, from the color
dependence of this systematic effect [39, 40].

We considered several sub-samples of HSD candidates. First, Geier
et al. [25] noted that some 10\% of the selected candidates could
represent be noise. Thus, it would be interesting to understand
how the kinematic parameters of the HSDs vary with position in the
HR diagram. Second, a dependence of the kinematic parameters of
the HSD candidates on Galactic latitude was detected in [26]. It
would be more correct to look for a dependence of these parameters
on distance from the Galactic $z.$ Accordingly, we compiled
several HSD samples at different distances from the Galactic
plane. Finally, the quality of the derived kinematic parameters
depends strongly on the relative uncertainties of the parallaxes
of the sample stars, $\sigma_\pi/\pi$. For this reason, we
consider here samples with different uncertainties,
$\sigma_\pi/\pi=30\%$ and $15\%.$

Figure 1 shows the HR diagram for the sample of HSDs. The absolute
magnitudes $M_G$ and color indices $G_{BP-RP}$ ($Gaia_{BP-RP}$)
were taken from [25]. All candidate HSDs with relative parallax
uncertainties below 15\% are plotted. We divided this sample into
four sections, labeled using Roman numbers. Figure 1a shows stars
with $|z|<0.5$ kpc and Fig. 1b those with $|z|\geq0.5$ kpc. In
addition, the stars in each panel are divided into two groups with
approximately the same number of stars, according to their
positions in the diagram. These two groups were identified as
follows. For color indices $G_{BP-RP}<0^m$, the boundary line
between the groups is given by
$M_G=\tan(80^\circ)*G_{BP-RP}+4^m.2$. The second boundary is the
vertical line at $G_{BP-RP}=0^m.$ We selected the two boundary
lines empirically, aiming to have approximately the same number of
stars in both samples without disrupting the general character of
the initial distribution.

 \begin{table}[t]
 \caption[]{\small
Parameters of the Galactic rotation derived from stars with
relative parallax uncertainties below 15\%
 }
  \begin{center}  \label{t:15}  \small
  \begin{tabular}{|l|r|r|r|r|r|}\hline
    Parameters                   &    All stars    &  $|z|<0.5$ kpc  & $|z|\geq0.5$ kpc \\\hline
      $N_\star$                  &            7766 &            4680 &            3086 \\
      ${\overline r},$ kpc       &            1.32 &            1.07 &            1.72 \\
  $|{\overline z}|,$ kpc         &          $0.50$ &         $ 0.24$ &          $0.81$ \\
    $U_\odot,$    km/s           &  $11.80\pm0.55$ &  $ 9.42\pm0.56$ &    $14.4\pm1.1$ \\
    $V_\odot,$    km/s           &  $35.00\pm0.80$ &  $25.44\pm0.84$ &    $52.6\pm1.6$ \\
    $W_\odot,$    km/s           &  $ 6.02\pm0.50$ &  $ 7.31\pm0.45$ &    $ 4.2\pm1.2$ \\
   $\Omega_0,$    km s$^{-1}$ kpc$^{-1}$ & $-26.78\pm0.50$ & $-28.68\pm0.58$ & $-23.89\pm0.90$ \\
  $\Omega^{'}_0,$ km s$^{-1}$ kpc$^{-2}$ & $  2.67\pm0.12$ & $  3.30\pm0.14$ & $  2.45\pm0.22$ \\
 $\Omega^{''}_0,$ km s$^{-1}$ kpc$^{-3}$ & $  0.43\pm0.16$ & $ -0.07\pm0.20$ & $  0.19\pm0.28$ \\
      $\sigma_0,$ km/s           &            38.2 &            29.2 &            52.0 \\
              $A$ km s$^{-1}$ kpc$^{-1}$       & $ 10.67\pm0.49$ & $ 13.20\pm0.57$ & $  9.80\pm0.90$ \\
              $B$ km s$^{-1}$ kpc$^{-1}$       & $-16.11\pm0.70$ & $-15.48\pm0.81$ & $-14.09\pm1.27$ \\
            $V_0$ km/s           & $   214\pm6$    & $   229\pm6$    & $   191\pm8$    \\\hline
      $\sigma_1,$ km/s           & $44.81\pm0.70$ & $36.74\pm0.53$  & $55.22\pm1.29$ \\
      $\sigma_2,$ km/s           & $39.20\pm1.10$ & $27.23\pm0.82$  & $50.39\pm1.82$ \\
      $\sigma_3,$ km/s           & $27.54\pm0.66$ & $22.40\pm0.74$  & $36.00\pm0.98$ \\
      $L_1, B_1$     & $~11\pm1^\circ,$\quad$~~~~0\pm1^\circ$ & $~~7\pm9^\circ,$\quad$~2\pm3^\circ$ & $~14\pm6^\circ,~$\quad$~-2\pm1^\circ$ \\
      $L_2, B_2$     & $101\pm6^\circ,$\quad$-10\pm2^\circ$   & $~97\pm3^\circ,$\quad$~5\pm2^\circ$ & $104\pm12^\circ,$\quad$-10\pm2^\circ$ \\
      $L_3, B_3$     & $100\pm6^\circ,$\quad$~~80\pm3^\circ$  & $254\pm3^\circ,$\quad$84\pm4^\circ$ & $~93\pm12^\circ,$\quad$~~80\pm4^\circ$ \\
  \hline
 \end{tabular}\end{center}
 {\small
$N_\star$ is the number of stars used, ${\overline r}$ the mean
distance of the stellar sample, and $\sigma_0$ the unit-weight
uncertainty. The lower part of the table presents the main axes
and orientations of the main axes for the residual velocity
ellipsoids of the corresponding samples.}
  \end{table}
 \begin{table}[t]
 \caption[]{\small
Parameters of the Galactic rotation derived from stars with
relative parallax uncertainties below 30\%
 }
  \begin{center}  \label{t:30}  \small
  \begin{tabular}{|l|r|r|r|r|r|}\hline
    Parameters                   &    All stars     &   $|z|<0.5$ kpc  & $|z|\geq0.5$ kpc \\\hline
      $N_\star$                  &            13253 &             6395 &             6858 \\
      ${\overline r},$ kpc       &             1.81 &             1.31 &             2.28 \\
  $|{\overline z}|,$ kpc         &           $0.69$ &          $ 0.35$ &           $1.08$ \\
    $U_\odot,$    km/s           &  $12.41\pm0.48$  &  $ 9.84\pm0.49$  &   $13.18\pm0.80$ \\
    $V_\odot,$    km/s           &  $39.96\pm0.66$  &  $25.81\pm0.70$  &   $58.12\pm1.15$ \\
    $W_\odot,$    km/s           &  $ 6.05\pm0.44$  &  $ 7.11\pm0.39$  &   $ 5.25\pm0.85$ \\
   $\Omega_0,$    km s$^{-1}$ kpc$^{-1}$ & $-24.69\pm0.31$  & $-28.40\pm0.39$  &  $-21.85\pm0.46$ \\
  $\Omega^{'}_0,$ km s$^{-1}$ kpc$^{-2}$ & $  2.53\pm0.08$  &  $ 3.46\pm0.09$  &   $ 2.29\pm0.12$ \\
 $\Omega^{''}_0,$ km s$^{-1}$ kpc$^{-3}$ & $ -0.04\pm0.06$  &  $-0.55\pm0.10$  &   $-0.29\pm0.09$ \\
      $\sigma_0,$ km/s           &             42.7 &             29.0 &             53.7 \\
              $A$ km s$^{-1}$ kpc$^{-1}$       &  $ 10.12\pm0.31$ &  $ 13.83\pm0.37$ &  $  9.16\pm0.46$ \\
              $B$ km s$^{-1}$ kpc$^{-1}$      &  $-14.57\pm0.43$ &  $-14.57\pm0.54$ &  $-12.68\pm0.65$ \\
            $V_0$ km/s           &  $   198\pm5$    &  $   227\pm5$    &  $   175\pm5$    \\\hline
      $\sigma_1,$ km/s           & $47.30\pm0.57$ &  $36.08\pm0.44$  &   $56.91\pm0.86$ \\
      $\sigma_2,$ km/s           & $45.65\pm0.84$ &  $27.55\pm0.76$  &   $55.76\pm1.12$ \\
      $\sigma_3,$ km/s           & $31.29\pm0.57$ &  $22.83\pm0.59$  &   $39.67\pm0.76$ \\
      $L_1, B_1$     & $~41\pm6^\circ,~$\quad$-4\pm1^\circ$  & $~10\pm7^\circ,$\quad$~2\pm2^\circ$ & $~37\pm8^\circ,~$\quad$-4\pm1^\circ$ \\
      $L_2, B_2$     & $131\pm14^\circ,$\quad$-6\pm1^\circ$  & $100\pm2^\circ,$\quad$~5\pm2^\circ$ & $128\pm19^\circ,$\quad$-3\pm2^\circ$ \\
      $L_3, B_3$     & $~96\pm14^\circ,$\quad$~83\pm2^\circ$ & $254\pm2^\circ,$\quad$85\pm3^\circ$ & $~73\pm19^\circ,$\quad$~85\pm2^\circ$ \\
  \hline
 \end{tabular}\end{center} {\small The notation in the Table is the same as in Table 1.}\end{table}
 \begin{table}[t]
 \caption[]{\small
Parameters of the Galactic rotation derived taking into account
the Lutz–Kelker bias
 }
  \begin{center}  \label{t:50}  \small
  \begin{tabular}{|l|r|r|r|r|r|}\hline
    Parameters             & $\sigma_\pi/\pi<15\%$ & $\sigma_\pi/\pi<30\%$   \\\hline
      $N_\star$                  &            7656 &           14415    \\
      ${\overline r},$ kpc       &            1.29 &            1.55    \\
    $U_\odot,$    km/s           &  $11.56\pm0.53$ &  $10.66\pm0.40$    \\
    $V_\odot,$    km/s           &  $34.50\pm0.80$ &  $33.52\pm0.55$    \\
    $W_\odot,$    km/s           &  $ 5.90\pm0.48$ &  $ 5.36\pm0.37$    \\
   $\Omega_0,$    km s$^{-1}$ kpc$^{-1}$ & $-26.82\pm0.50$ & $-24.87\pm0.31$    \\
  $\Omega^{'}_0,$ km s$^{-1}$ kpc$^{-2}$ & $  2.69\pm0.12$ &  $ 2.44\pm0.07$    \\
 $\Omega^{''}_0,$ km s$^{-1}$ kpc$^{-3}$ & $  0.38\pm0.17$ &  $ 0.38\pm0.07$    \\
      $\sigma_0,$ km/s                   &            37.3 &             37.7   \\
              $A$ km s$^{-1}$ kpc$^{-1}$ & $ 10.77\pm0.48$ &  $  9.74\pm0.30$   \\
              $B$ km s$^{-1}$ kpc$^{-1}$ & $-16.06\pm0.68$ &  $-15.13\pm0.43$   \\
              $V_0$ km/s         & $   215\pm6$    &  $   199\pm5$      \\\hline
      $\sigma_1,$ km/s           &  $42.35\pm0.59$ &  $42.52\pm0.47$    \\
      $\sigma_2,$ km/s           &  $36.15\pm0.77$ &  $40.07\pm0.76$    \\
      $\sigma_3,$ km/s           &  $26.55\pm0.48$ &  $27.73\pm0.46$    \\
      $L_1, B_1$     & $~9\pm1^\circ,$ ~\quad$~0\pm1^\circ$ & $~23\pm4^\circ, $ \quad$-2\pm1^\circ$ \\
      $L_2, B_2$     & $99\pm4^\circ,$  \quad$-6\pm1^\circ$ & $114\pm12^\circ,$ \quad$-7\pm1^\circ$ \\
      $L_3, B_3$     & $99\pm4^\circ,$ ~\quad$84\pm2^\circ$ & $~96\pm12^\circ,$ \quad$83\pm2^\circ$ \\
  \hline
 \end{tabular}\end{center} {\small The notation in the Table is the same as in Table 1.}\end{table}
\begin{figure}[t]
{\begin{center}
   \includegraphics[width=0.7\textwidth]{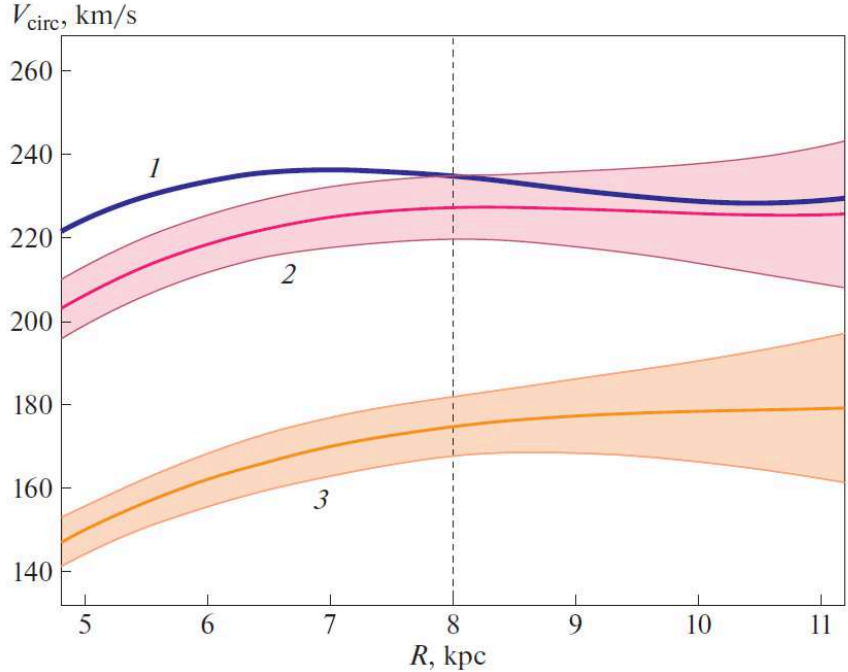}
 \caption{
Galactic rotation curves plotted according to [15] (curve 1), the
solution in the second-to-last column in Table 2 (curve 2), and
the solution in the last column of Table 2 (curve 3). The
boundaries of the confidence intervals correspond to $1\sigma.$
The vertical dashed line shows the position of the Sun.
  } \label{f-Rotat}
\end{center}}
\end{figure}

 \section*{RESULTS AND DISCUSSION}
 \subsection*{Galactic Rotation Parameters}
We first divided the entire sample into groups in terms of the
absolute values of their $z$ coordinates, $|z|.$ We also performed
a division based on the relative uncertainty of their parallaxes.
Using stars with relative parallax uncertainties below 30\%
provides a large number of objects and enables us to estimate the
parameters of interest with lower errors. On the other hand, the
sample with parallax uncertainties below 15\% enables us to obtain
more local parameters, in particular, more reliable estimates of
the $(U,V,W)_\odot$ and $\Omega_0$.

The kinematic parameters found for these stars are collected in
Tables 1 and 2. In addition to the six desired kinematic
parameters, they contain the the mean distance of the sample stars
${\overline r},$ mean $|z|,$ unit-weight uncertainty $\sigma_0,$
estimated from a least squares solution of the system of
conditional equations of the form (1)--(2), and the Oort constants
$A$ and $B,$ calculated from the relations
\begin{equation}
 A= 0.5\Omega^{'}_0R_0,\quad
 B=-\Omega_0+A,
 \label{AB}
\end{equation}
The linear velocity of the Galactic rotation in the solar vicinity
$V_0=|R_0\Omega_0|$ is also given. Note that we calculated the
kinematic parameters presented in Tables 1 and 2 using stars with
$r<6$ kpc and rejected stars with large (above 40 km/s) random
errors in their measured proper motions.

The derived velocities $(U,V,W)_\odot$ and the Oort constants
$A,B$ can be used to correct the velocities $V_l,$ and $V_b$:
 \begin{equation}
 \begin{array}{lll}
 V_l^*= V_l-(U_\odot\sin l-V_\odot\cos l)
         -(A\cos 2l+B)r\cos b,\\
 V_b^*= V_b-(U_\odot\cos l\sin b+V_\odot\sin l\sin b
         -W_\odot\cos b)+Ar\sin 2l\sin b\cos b.
 \label{EQ-888}
 \end{array}
 \end{equation}
After this procedure, the residual velocities $V_l^*$ and $V_b^*$
should be used on the left-hand sides of (7)--(9). On the other
hand, we can also form the residual velocities with the derived
values of $\Omega_0$, $\Omega^{'}_0$ and $\Omega^{''}_0;$ note
that the last term can be disregarded, since its influence in the
region we are considering is negligible.

 \subsubsection*{The Influence of the Lutz–Kelker Bias}
Using the inverses of the parallaxes when estimating the distances
results in a systematic bias of the distances obtained [27,
41--43]. The magnitude of this systematic bias depends
significantly on the relative accuracy of the parallaxes and
increases with the distance.

We calculated the Galactic rotation parameters and the parameters
of the residual velocity ellipsoids for the two star samples
taking into account the Lutz–Kelker bias. For each star with an
input $\sigma_\pi/\pi$ value, we calculated the $G(Z)$
distribution functions using the formula modified in [27] for the
case of objects with a flat space distribution:
\begin{equation}
\label{LK_equation}
 G(Z)\sim Z^{-3}\exp\bigg{[}-\frac{(Z-1)^2}{2(\sigma_\pi/\pi)^2}\bigg{]},
\end{equation}
where $Z=\pi/\pi_{true}$. The $G(Z)$ distribution function was
then used to determine the factor $Z$ and the new distance,
$r_{true}=1/\pi_{true}$ (the result is that the initial distances
became smaller).

For this purpose, we selected HSDs with relative parallax
uncertainties below 15\% (the ``All stars'' column in Table 1) and
below 30\% (the ``All stars'' column in Table 2). The results are
presented in Table 3.

These calculations demonstrated the following. For stars with
relative parallax uncertainties below 15\%, taking into account
the influence of the Lutz--Kelker bias on the velocities of the
solar motion and the Galactic rotation results in only a
negligible change in the determined parameters. This influence is
slightly more significant, though still small, for the dispersions
of the relative velocities: $\sigma_1,\sigma_2,\sigma_3$. In
general, our results agree with the conclusion of Lutz and Kelker
[41] that the critical parallax error (below which there is no
need to correct for the bias) is $\sigma_\pi/\pi:15\%-20\%$. This
is also in agreement with the conclusions of [42], where it was
proposed to consider a sample with parallaxes from the Gaia DR2
catalog ``good'' if the uncertainties for individual stars were
$\sigma_\pi/\pi<20\%$. This led us to disregard corrections for
the Lutz--Kelker bias in our analysis of stars with uncertainties
$\sigma_\pi/\pi<15\%$.

Comparing the corresponding parameters presented in Tables 2 and
3, we can see that taking into account the Lutz–Kelker bias for
stars with uncertainties $\sigma_\pi/\pi<30\%$ results in a more
appreciable change in i the kinematic parameters. In this case,
this correction has a favorable influence on the results derived
from these stars. In particular, the velocity dispersions and
alignment parameters of the velocity ellipsoids became closer for
the two star samples presented in Table 3 after taking into
account the correction.

 \subsubsection*{Estimating the gradient $\Delta V_0/\Delta |z|$}
Figure 2 shows three Galactic rotation curves. The first curve was
computed by Bobylev and Bajkova [15] for a sample of stars in
young open clusters. The second curve corresponds to the solution
presented in the next-to-last column in Table 2, and the third
curve to the solution in the last column of Table 2. The
difference in the Galactic rotation velocities near the solar
distance for the two derived curves is $\Delta V_0=52$ km/s.

We can use the data from Tables 1 and 2 to estimate the gradient
of the circular rotation velocity, $V_0,$ in $|z|.$ We found that
${\displaystyle\Delta V_0\over\displaystyle \Delta |z|}=-67\pm10$
km s$^{-1}$ kpc$^{-1}$ for the sample of stars with relative
parallax uncertainties $<15\%$ and ${\displaystyle\Delta
V_0\over\displaystyle \Delta |z|}=-71\pm7$ km s$^{-1}$ kpc$^{-1}$
for somewhat more distant and ``higher'' stars with relative
parallax uncertainties $<30\%.$ Chiba and Beers [44] obtained the
following estimates of this gradient for a sample of low
metallicity stars in the solar neighborhood: ${\displaystyle\Delta
V_0\over\displaystyle \Delta |z|}=-30\pm3$ km s$^{-1}$ kpc$^{-1}$
for thin-disk stars with the velocity ellipsoid
$(\sigma_U,\sigma_V,\sigma_W)=(46,50,35)\pm(4,4,3)$ km/s;
${\displaystyle\Delta V_0\over\displaystyle \Delta |z|}=-52\pm6$
km s$^{-1}$ kpc$^{-1}$ for halo stars with a very elongated
velocity ellipsoid,
$(\sigma_U,\sigma_V,\sigma_W)=(141,106,94)\pm(11,9,8)$ km/s.

 \begin{table}[t]
 \caption[]{\small
Parameters of the residual-velocity ellipsoids for the four
samples of HSDs located in four $z$ zones. Stars with relative
parallax uncertainties below 15\% were used
 }
  \begin{center}  \label{t:3}   \small
  \begin{tabular}{|l|r|r|r|r|r|}\hline
  Parameters           &  $z<-0.5$ kpc & $-0.5\leq z<0$ kpc & $0\leq z<0.5$ kpc & $z\geq0.5$ kpc \\\hline
  $N_\star$            &          1563 &          2494 &           2185 &             1523 \\
  ${\overline r},$ kpc &          1.71 &          1.07 &           1.06 &             1.75 \\
  ${\overline z},$ kpc &       $-0.91$ &       $-0.23$ &         $0.22$ &           $0.93$ \\
  $\sigma_1,$ km/s     & $ 57.1\pm1.8$ &  $36.3\pm0.9$ &   $36.9\pm0.6$ &   $55.9\pm1.7$ \\
  $\sigma_2,$ km/s     & $ 52.0\pm2.6$ &  $28.5\pm1.1$ &   $26.9\pm1.2$ &   $49.1\pm2.6$ \\
  $\sigma_3,$ km/s     & $ 30.7\pm1.8$ &  $22.3\pm1.3$ &   $21.6\pm0.7$ &   $34.3\pm1.3$ \\
  $L_1, B_1$  & $~~1^\circ,$ $~~12\pm4^\circ$ & $~~7\pm4^\circ,$ $~~6\pm5^\circ$ & $~11\pm2^\circ,$ $~~0\pm1^\circ$ & $~10^\circ,$ $-13\pm4^\circ$ \\
  $L_2, B_2$  & $~89^\circ,$  $-10\pm4^\circ$ & $~97\pm4^\circ,$ $~~0\pm4^\circ$ & $101\pm3^\circ,$  $10\pm2^\circ$ & $101^\circ,$ $-~5\pm3^\circ$ \\
  $L_3, B_3$  & $140^\circ,$ $~~74\pm4^\circ$ & $188\pm4^\circ,$  $84\pm5^\circ$ & $283\pm3^\circ,$  $80\pm5^\circ$ & $~32^\circ,$ $~~76\pm5^\circ$ \\
  \hline
  \end{tabular}\end{center} \end{table}
\begin{figure}[t]
{\begin{center}
   \includegraphics[width=0.5\textwidth]{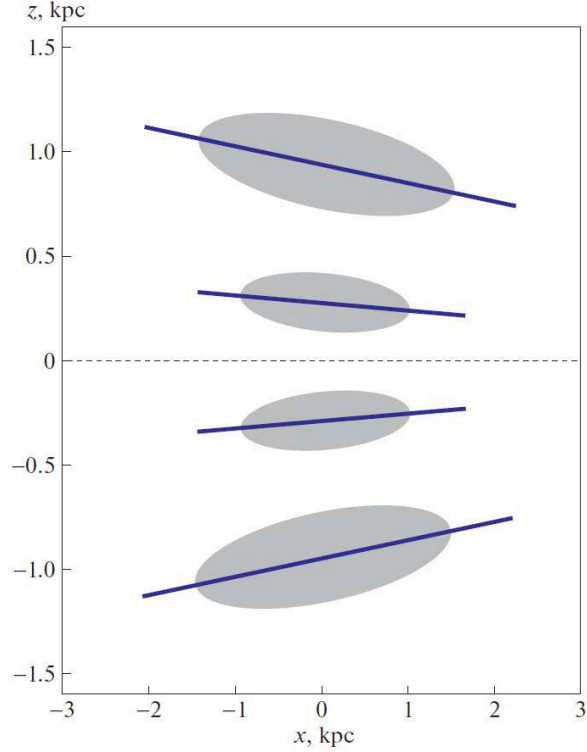}
 \caption{
Schematic representation of the four derived ellipses for the $UW$
residual velocities on the $xz$ Galactic plane.
  } \label{f-ellipses}
\end{center}}
\end{figure}

 \subsection*{Parameters of the Velocity Ellipsoid}
Table 4 presents the parameters of the residual velocity
ellipsoids for the four samples of HSDs in the four z zones, two
above and two below the Galactic plane. The boundaries were chosen
to provide approximately the same numbers of stars in each of the
samples. We took into account the dependence of the Galactic
rotation on $|z|;$ i.e., we used the two rotation curves displayed
in Fig. 2. The position of the first axis $L_1, B_1$ deserves
special attention, and is confirmed by the independently derived
position of the third axis.

In the first column of Table 4, $B_1=12\pm4^\circ$ for the most
negative $z$ values, while $B_1=-13\pm4^\circ$ for the most
positive values of $z$ in the last column. The directions L1 are
close, though they were determined in these zones with large
uncertainties (and, for this reason, are not given in the table).
Thus, we have two ellipses in the $UW$ plane, located
symmetrically about the Galactic plane at heights ${\overline
z}=\pm0.9$ kpc; their first axes are directed towards the Galactic
center at the same angles, $\mp12^\circ.$ Figure 3 shows (not to
scale) the projections of the four derived ellipses of the $UW$
residual velocities onto the Galactic plane $xz$.

Note also that the third axis is closer to the direction towards
the pole in two regions adjacent to the Galactic plane ($|z|<0.5$
kpc), compared to regions located higher. There probably also
exist effects from inhomogeneities in other planes, since some
$B_2$ values differ significantly from zero.

 \begin{table}[t]
 \caption[]{\small
Kinematic parameters of the star samples with relative parallax
uncertainties below 15\%, located in two $|z|$ zones
 }
  \begin{center}  \label{t:04}  \small
  \begin{tabular}{|l|r|r|r|r|r|}\hline
  Parameters           &             I &            II &           III &          IV  \\
                       & $|z|<0.5$~kpc & $|z|<0.5$~kpc & $|z|\geq0.5$~kpc & $|z|\geq0.5$~kpc \\\hline
  $N_\star$            &          2733 &          1947 &          2049 &         1037 \\
  ${\overline r},$ kpc &          1.08 &          1.04 &          1.55 &         2.05 \\
  ${\overline z},$ kpc &       $-0.91$ &       $-0.23$ &        $0.22$ &       $0.93$ \\
  $U_\odot,$ km/s     & $10.40\pm0.76$ & $ 7.64\pm0.83$ & $13.7\pm1.1$ & $11.5\pm2.5$ \\
  $V_\odot,$ km/s     & $28.98\pm1.25$ & $20.97\pm1.14$ & $49.2\pm1.8$ & $72.7\pm3.8$ \\
  $W_\odot,$ km/s     & $ 8.15\pm0.64$ & $ 6.12\pm0.63$ & $ 5.1\pm1.3$ & $ 4.3\pm2.7$ \\
   $\Omega_0,$    km s$^{-1}$ kpc$^{-1}$ & $-28.23\pm0.78$ & $-29.44\pm0.85$ & $-24.48\pm1.05$ & $-21.2\pm1.7$  \\
  $\Omega^{'}_0,$ km s$^{-1}$ kpc$^{-2}$ & $  3.43\pm0.20$ &  $ 3.18\pm0.20$ &  $ 2.86\pm0.26$ & $ 1.89\pm0.41$ \\
 $\Omega^{''}_0,$ km s$^{-1}$ kpc$^{-3}$ & $ -0.53\pm0.34$ &  $ 0.37\pm0.23$ &  $-1.17\pm0.41$ & $-0.23\pm0.46$ \\
  $\sigma_0,$ km/s    & $ 30.8$      & $ 26.7$         & $ 43.9      $ & $ 65.1$       \\
  $A,$   km s$^{-1}$ kpc$^{-1}$  & $ 13.73\pm0.80$ & $ 12.73\pm0.80$ & $ 11.4\pm1.1$ & $  7.6\pm1.6$ \\
  $B,$   km s$^{-1}$ kpc$^{-1}$  & $-14.50\pm1.11$ & $-16.71\pm1.17$ & $-13.1\pm1.5$ & $-13.7\pm2.3$ \\
  $V_0,$ km/s      & $   226\pm8   $ & $   236\pm8   $ & $  196\pm9  $ & $  170\pm14 $ \\\hline
  $\sigma_1,$ km/s     & $ 35.84\pm0.45$ &  $31.02\pm0.72$ &   $46.5\pm1.2$ &   $67.5\pm1.9$ \\
  $\sigma_2,$ km/s     & $ 26.96\pm0.55$ &  $25.08\pm0.84$ &   $39.6\pm1.3$ &   $61.3\pm2.1$ \\
  $\sigma_3,$ km/s     & $ 22.29\pm0.41$ &  $19.21\pm0.56$ &   $32.1\pm0.8$ &   $41.8\pm1.5$ \\
  $L_1, B_1$  & $~~6\pm8^\circ,$ $~~2\pm4^\circ$ & $~15\pm9^\circ,$ $~3\pm3^\circ$ & $~7\pm1^\circ,$ $~~0\pm1^\circ$ & $-13\pm7^\circ,$ $~~2\pm1^\circ$ \\
  $L_2, B_2$  & $~96\pm2^\circ,$  $-1\pm4^\circ$ & $106\pm5^\circ,$ $~5\pm2^\circ$ & $98\pm7^\circ,$  $-7\pm3^\circ$ & $77\pm14^\circ,$  $-8\pm3^\circ$ \\
  $L_3, B_3$  & $156\pm2^\circ,$  $87\pm4^\circ$ & $252\pm5^\circ,$ $84\pm4^\circ$ & $95\pm7^\circ,$  $83\pm6^\circ$ & $90\pm14^\circ,$  $82\pm4^\circ$ \\
  \hline
  \end{tabular}\end{center}
  {\small The notation in the Table is the same as in Table 1.}\end{table}

Anguiano et al. [45] found the following dispersions from their
analysis of stellar proper motions and parallaxes from the Gaia
DR1 catalog [3]:
 $(\sigma_U,\sigma_V,\sigma_W)=(33,28,23)\pm(4,2,2)$ km/s for
thin-disk stars and
 $(\sigma_U,\sigma_V,\sigma_W)=(57,38,37)\pm(6,5,4)$ km/s
for thick-disk stars. They demonstrated that the deviation of the
vertex in the $UV$ plane for different stellar groups varied in a
very wide range, from $-5^\circ$ to $+40^\circ$, while the
inclination in the $UW$ plane varied from $-10^\circ$ to
$+15^\circ$ Tables 1 and 2 show that the $z$ division we used
enables us to identify stars with properties close to the
kinematics of the thin and thick disks.

Bobylev and Bajkova [26] found the following dispersions for a
sample of low-latitude HSDs with parallax uncertainties below
15\%:
$(\sigma_1,\sigma_2,\sigma_3)=(37.4,28.1,22.8)\pm(0.9,0.7,0.9)$
km/s. The dispersions for a high-latitude sample were
$(\sigma_1,\sigma_2,\sigma_3)=(51.9,46.6,34.8)\pm(1.1,1.8,0.8)$
km/s. We can see an agreement with the results of the analysis of
the same stars presented in Table 1, though $\sigma_1$ and
$\sigma_2$ for the high-latitude HSDs were found to be lower than
the corresponding values for our subdivision into $|z|$ layers
(the last column of Table 1). This difference is due primarily to
the boundaries used when forming the samples.

Using data from the Gaia DR2 catalog, Hagen et al. [46] performed
a large-scale study of the $(V_R, V_z)$ velocity plane with
respect to the spatial positions of stars. They considered a
region with a radius of about 4 kpc around the Sun, using stars
from the Gaia DR2 catalog. Evolution of the size and alignment of
the residual velocity ellipsoid with $R$ and $z$ was demonstrated.
Namely, they plotted maps with a large number of ellipses whose
first axes was nearly always directed towards the Galactic center
(far outside the solar circle, at $R\sim12$ kpc, the inclination
becomes close to zero for any $z$). In this sense, our Fig. 3 is
in agreement with the results of Hagen et al. [46], while each
study derived its own sizes for the ellipses for each of the
Galactic subsystems.

Examining Fig. 1, we have the impression that there are a fair
number of main-sequence stars among the HSD candidates. We
accordingly decided to trace changes in the kinematic parameters
for the samples I--IV. Table 5 presents the kinematic parameters
for four samples identified in accordance with the samples labeled
with Roman numbers I--IV in Fig. 1.

We can see from Fig. 1 that the amount of contamination is small
for samples I and III. This is especially true for sample III,
where we see a well expressed, virtually isolated cluster of stars
just in the region where HSDs should be located. The same is true
for sample I; however, here, on the contrary, some of the
``hundred-percent'' HSDs are cut off and appear in sample II. This
happened because the main clump of HSDs in Fig. 1a is more
extended along the $G_{BP-RP}$ coordinate, compared to Fig. 1b.
Thus, sample II also contains a large ($>60\%$) fraction of HSDs.
The greatest problems in this respect occur for the sample IV,
where we have many stars in the main-sequence region that are
clearly separated from the main clump III. As follows from [25,
Fig. 2], regions II and IV are dominated by binaries containing a
cool main-sequence star as one of their components. The number of
such systems is apparently fairly high in region IV.

Table 5 shows that sample II seems to be the kinematically
youngest: here, the stars show the highest rotation velocity,
$V_0,$ and the lowest residual-velocity dispersions,
$\sigma_1,\sigma_2,\sigma_3.$ The highest kinematic age is
displayed by stars from sample IV: they have the lowest rotation
velocity, $V_0,$ and the highest residual velocity dispersions,
$\sigma_1,\sigma_2,\sigma_3.$ Note that the parameters of the
Galactic rotation curve derived from sample I are in good
agreement with known results (and with those discussed by us
above), including the second derivative of the rotation angular
velocity, $\Omega^{''}_0.$

The velocity $V_\odot\sim21$ km/s derived for stars in sample II
demonstrates that they lag behind the LSR by only $V_{lag}=\Delta
V_\odot\sim8$ km/s, due to the so-called asymmetric drift. Here,
we used one of the most reliable modern determinations of the
Sun's peculiar motion relative to the LSR [47]:
$(U_\odot,V_\odot,W_\odot)=(11.1,12.2,7.3)\pm(0.7,0.5,0.4)$ km/s.
The asymmetric drift increases the $V_{lag}$ velocity for all old
Galactic objects. It follows from the last column of Table 5 that
the lag of the sample IV stars behind the LSR is considerable,
$V_{lag}\sim60$ km/s. On the other hand, we can see from Tables 1,
2, and 5 that there is no significant deviation from the standard
value for the $U_\odot$ velocities; a slight difference in the
$W_\odot$ velocity is seen for stars at high $z.$

We can use the ``hundred-percent'' HSDs, i.e., samples II and III,
to estimate the gradient of the circular rotation velocity, $V_0,$
in $|z|.$ This gradient is ${\displaystyle\Delta
V_0\over\displaystyle \Delta |z|}=-52\pm11$ km s$^{-1}$ kpc$^{-1}$
— fairly moderate, closer to the more reliable results obtained by
Chiba and Beers [44] based on a large number of data points.
Samples I and IV give the estimate ${\displaystyle\Delta
V_0\over\displaystyle \Delta |z|}=-74\pm16$ km s$^{-1}$
kpc$^{-1}$.

The parameters of the residual stellar-velocity ellipsoids
presented in Table 5 agree with the parameters in Table 1. Since
the sign of the velocity is averaged when considering $|z|$
layers, we do not find any serious deviations in the ellipsoid
alignment with respect to the $x, y, z$ coordinate axes. Though
the semi-major axes of the velocity ellipsoid found for sample IV
are the largest among those derived in our study, they
nevertheless fall short of corresponding values for the halo.

 \section*{CONCLUSION}
We have studied the kinematics of HSDs from the catalog of Geier
et al. [25], selected from the Gaia DR2 catalog, in conjunction
with data from several multi-band photometric sky surveys. Our
study made use of more than 13 000 proper motions for stars with
relative parallax uncertainties below 30\%. A zero-point
correction, $\Delta\pi=0.050$ mas, was added to all parallaxes
from the Gaia DR2 catalog.

We used two samples of stars with relative parallax uncertainties
below 15\% and 30\% to study the influence of the Lutz–Kelker bias
[41]. The influence of this bias on the derived parameters of the
Sun's peculiar motion, Galactic rotation, and residual-velocity
ellipsoid was demonstrated to be negligible for uncertainties
below 15

We determined the parameters of the Galactic rotation for two
$|z|$ zones. The linear rotation velocity found for the HSDs at
$|z|<0.5$ kpc is $V_0=227\pm5$ km/s. This indicates that these
stars belong to the Galactic thin disk, as is also confirmed by
the semi-major axes of their residual velocity ellipsoid:
$(\sigma_1,\sigma_2,\sigma_3)=(36.1,27.6,22.8)\pm(0.4,0.8,0.6)$
km/s.

The HSDs at $|z|\geq0.5$ kpc display a considerably lower rotation
velocity, $V_0=175\pm5$ km/s, typical of thick-disk objects. The
semi-major axes found for their residual-velocity ellipsoid also
suggest membership in the thick disk:
$(\sigma_1,\sigma_2,\sigma_3)=(56.9,55.8,39.7)\pm(0.9,1.1,0.8)$
km/s. We used these data to estimate the gradient of the circular
rotation velocity, $V_0,$ in $|z|$ to be $\Delta V_0/\Delta
|z|=-71\pm7$ km s$^{-1}$ kpc$^{-1}$.

We also considered samples with the most probable HSD candidates.
These give the estimate $\Delta V_0/\Delta |z|=-52\pm11$ km
s$^{-1}$ kpc$^{-1}$. Using such stars (sample III) at $|z|\geq0.5$
kpc zone, we found more moderate semi-major axes for their
residual velocity ellipsoid:
$(\sigma_1,\sigma_2,\sigma_3)=(46.5,39.6,32.1)\pm(1.2,1.3,0.8)$
km/s.

When forming the stellar residual velocities, we took into account
the Galactic rotation separately for each $z$ zone. The parameters
of the residual velocity ellipsoids were obtained for HSDs in four
plane-parallel layers. The the size of the ellipsoid increases
with $z,$ as does the inclination of the first axis to the
Galactic plane.

We have shown that, at $|z|<0.5$ kpc, the possible contamination
of the sample of candidate HSDs comprises only a small fraction,
related to kinematically cooler, younger main-sequence stars that
have little influence on the derived kinematic parameters.
However, the contamination could be higher at $|z|\geq0.5$ kpc,
associated with stars with higher velocity dispersions; i.e.,
stars that are hotter in a kinematic sense.

 \subsubsection*{ACKNOWLEDGEMENTS}
The authors thank the referee for useful remarks that helped us to
improve the paper.

 \medskip\subsubsection*{REFERENCES}

 {\small
 \quad ~ 1. M.L. Humason and F. Zwicky, Astrophys. J. {\bf 105}, 85 (1947).

 2. J.L. Greenstein and A.I. Sargent, Astrophys. J. Suppl. {\bf 28}, 157 (1974).

 3. Gaia Collaboration, A.G.A. Brown, A. Vallenari, T. Prusti, J. de Bruijne, F. Mignard,
    R. Drimmel, et al., Astron. Astrophys. {\bf 595}, 2 (2016). 

 4. Gaia Collaboration, A.G.A. Brown, A. Vallenari, T. Prusti, de Bruijne, C. Babusiaux,
    C.A.L. Bailer-Jones, M. Biermann, D.W. Evans, et al.,
    Astron. Astrophys. {\bf 616}, 1 (2018). 

 5. Gaia Collaboration, L. Lindegren, J. Hernandez, A. Bombrun, S. Klioner, U. Bastian,
    M. Ramos-Lerate, A. de Torres, H. Steidelmuller, et al.,
    Astron. Astrophys. {\bf 616}, 2 (2018). 

 6. G. Iorio and V. Belokurov, Mon. Not. R. Astron. Soc. {\bf 482}, 3868 (2019).

 7. N. Rowell and M. Kilic, Mon. Not. R. Astron. Soc. {\bf 484}, 3544 (2019).

 8. T. Antoja, A. Helmi, M. Romero-G\'omez, D. Katz, C. Babusiaux, R.
    Drimmel, D. W. Evans, F. Figueras, et al., Nature {\bf 561}, 360 (2018).

 9. M. Bennett and J. Bovy, Mon. Not. R. Astron. Soc. {\bf 482}, 1417 (2019).

 10. E. Vasiliev, Mon. Not. R. Astron. Soc. {\bf 484}, 2832 (2019).

 11. H. Baumgardt, M. Hilker, A. Sollima, and A. Bellini,
    Mon. Not. R. Astron. Soc. {\bf 482}, 5138 (2019).

 12. G. Eadie and M. Juric, Astrophys. J. {\bf 875}, 159 (2019).

 13. D. Kawata, J. Bovy, N. Matsunaga, and J. Baba, Mon. Not. R. Astron. Soc. {\bf 482}, 40 (2019).

 14. V.V. Bobylev and A.T. Bajkova, Astron. Lett. {\bf 44}, 675 (2018). 

 15. V.V. Bobylev and A.T. Bajkova, Astron. Lett. {\bf 45}, 109 (2019). 

 16. J.A.S. Hunt, J. Hong, J. Bovy, D. Kawata, and R.J.J. Grand,
     Mon. Not. R. Astron. Soc. {\bf 481}, 3794 (2018).

 17. J.A. Sellwood, W.H. Trick, R.G. Carlberg, J. Coronado, and H.-W. Rix,
     Mon. Not. R. Astron. Soc. {\bf 484}, 3154 (2019).

 18. W.S. Dias, H. Monteiro, J.R.D. L\'epine, R. Prates, C.D. Gneiding, and M. Sacchi,
     Mon. Not. R. Astron. Soc. {\bf 481}, 3887 (2018).

 19. C. Soubiran, T. Cantat-Gaudin, M. Romero-Gomez, L. Casamiquela, C. Jordi, A. Vallenari,
     T. Antoja, L. Balaguer-N\'u\~nez, et al., Astron. Astrophys. {\bf 619}, 155 (2018).

 20. M. Altmann, H. Edelmann, and K.S. de Boer, Astron. Astrophys. {\bf 414}, 181 (2004).

 21. S.K. Randall, S. Bagnulo, E. Ziegerer, S. Geier, and G. Fontaine,
     Astron. Astrophys. {\bf 576}, 65 (2015).

 22. P. Martin, C.S. Jeffery, N. Naslim, and V.M. Woolf, Mon. Not. R. Astron. Soc.
     {\bf 467}, 68 (2017).

 23. E.-M. Pauli, R. Napiwotzki, U. Heber, M. Altmann, and M. Odenkirchen,
     Astron. and Astrophys. {\bf 447}, 173 (2006).

 24. Y. Bu, Z. Lei, G. Zhao, J. Bu, and J. Pan, Astrophys. J. Suppl. Ser. {\bf 233}, 2 (2017).

 25. S. Geier, R. Raddi, N.P. Gentile Fusillo, and T.R. Marsh,
     Astron. Astrophys. {\bf 621}, 38 (2019).

 26. V.V. Bobylev and A.T. Bajkova, Astron. Lett. 45 (2019, in press). 

 27. A.S. Rastorguev, M.V. Zabolotskikh, A.K. Dambis, et al.,
     Astrophys. Bulletin, 72, 122 (2017).

 28. V.V. Vityazev, A.S. Tsvetkov, V.V. Bobylev, et al., Astrophysics, {\bf 60}, 462 (2017).

 29. V.V. Bobylev and A.T. Bajkova, Astron. Lett. {\bf 43}, 452 (2017). 

 30. J.P. Vall\'ee, Astrophysics and Space Science {\bf 362}, 79 (2017). 

 31. R. de Grijs and G. Bono, Astrophys. J. Suppl. Ser. {\bf 232}, 22 (2017). 

 32. T. Camarillo, M. Varun, M. Tyler, and R. Bharat), PASP {\bf 130}, 4101 (2018). 

 33. K.F. Ogorodnikov, {\it Dynamics of stellar systems} (Oxford: Pergamon, ed. Beer, A. 1965).

 34. Gaia Collaboration, F. Arenou, X. Luri, C. Babusiaux, C. Fabricius, A. Helmi,
    T. Muraveva, A.C. Robin, F. Spoto, et al., Astron. Astrophys. {\bf 616}, 17 (2018).

 35. K.G. Stassun and G. Torres, Astrophys. J. {\bf 862}, 61 (2018).

 36. A.G. Riess, S. Casertano, W. Yuan, L. Macri, B. Bucciarelli, M.G. Lattanzi,
     J.W. MacKenty, J.B. Bowers, et al., Astrophys. J. {\bf 861}, 126 (2018).

 37. J.C. Zinn, M.H. Pinsonneault, D. Huber, and D. Stello, arXiv: 1805.02650 (2018).

 38. L.N. Yalyalieva, A.A. Chemel, E.V. Glushkova, A.K. Dambis, and A.D. Klinichev,
     Astrophys. Bulletin, {\bf 73}, 335 (2018).

 39. H.W. Leung and J. Bovy, arXiv:1902.08634 (2019).

 40. D. Graczyk, G. Pietrzynski, W. Gieren, J. Storm, N. Nardetto, A. Gallenne, P. Maxted,
     P. Kervella, et al., Astrophys. J. {\bf 872}, 85 (2019).

 41. T.E. Lutz and D.H. Kelker, PASP {\bf 85}, 573 (1973).

 42. C.A.L. Bailer-Jones, PASP {\bf 127}, 994 (2015).

 43. Gaia Collaboration, X. Luri, A.G.A. Brown, L.M. Sarro, F. Arenou, C.A.L. Bailer-Jones,
      A. Castro-Ginard,  J. de Bruijne, T. Prusti, et al.,
      Astron. and Astrophys. {\bf 616}, 9 (2018).

 44. M. Chiba and T.C. Beers, Astron. J. {\bf 119}, 2843 (2000).

 45. B. Anguiano, S.R. Majewski, K.C. Freeman, A.W. Mitschang, and M.C. Smith,
     Mon. Not. R. Astron. Soc. {\bf 474}, 854 (2018).

 46. J.H.J. Hagen, A. Helmi, P.T. de Zeeuw, and L. Posti, arXiv:1902.05268 (2019).

 47. R. Sch\"onrich, J. Binney, and W. Dehnen,
     Mon. Not. R. Astron. Soc. {\bf 403}, 1829 (2010).

 }
 \end{document}